# Nonlinear modeling of the scaling law for the $m/n$=3/2 error field penetration threshold

**Q. Hu[1], N.C. Logan[1], J-K. Park[1], C. Paz-Soldan[2], R. Nazikian[1] and Q. Yu[3]**

[1]*Princeton Plasma Physics Laboratory, Princeton NJ 08543-0451, USA*

[2]*General Atomics, PO Box 85608 San Diego, CA 92186-5608, USA* [3]*Max-Plank-Institut fu¨r Plasmaphysik, 85748 Garching, Germany*

E-mail: qhu@pppl.gov

**Abstract.** The scaling law for the $m/n$ = 3/2 error field (EF) penetration threshold is predicted numerically based on nonlinear *single-fluid* and *two-fluid* modeling using the TM1 code. The simulated penetration threshold of radial magnetic field $b_r$ at the plasma edge is scaled to the electron density $n_e$, temperature $T_e$, viscous time $\tau_\mu$, toroidal field $B_t$ and the natural frequency $\omega$ in the form of $b_r/B_t \propto n_e^{\alpha_n} T_e^{\alpha_T} \tau_\mu^{\alpha_\mu} B_t^{\alpha_B} \omega^{\alpha_\omega}$ by scanning these parameters separately. Here, $\alpha_n$, $\alpha_T$, $\alpha_\mu$, $\alpha_B$ and $\alpha_\omega$ are the scaling coefficients on $n_e$, $T_e$, $\tau_\mu$, $B_t$ and $\omega$, respectively. Single fluid modeling shows that the 3/2 EF threshold scales as $b_r/B_t \propto n_e^{0.56} T_e^{0.6} \tau_\mu^{-0.59} B_t^{-1.15} \omega$, which is similar with the analytical scaling law in both the *Rutherford* and *visco-resistive* regimes. However, two-fluid modeling shows that the scaling law differs significantly in particular regarding the dependence on plasma rotation. In detail, the scaling coefficient $\alpha_n$ on density decreases from 0.67 to 0.56 and $\alpha_T$ on temperature decreases from 0.67 to 0.32, while $\alpha_\mu$ on viscous time is around -0.45 and $\alpha_B$ on toroidal field decreases slightly from -1.15 to -1, when the ratio $|\omega_E/\omega_{*e}|$ between plasma rotation frequency $\omega_E$ and diamagnetic drift frequency $\omega_{*e}$ varies from 0 to 10. Scans of the plasma rotation reveals that the penetration threshold linearly depends on the perpendicular rotation frequency (or natural frequency) $\omega_{\perp e} = \omega_E + \omega_{*e}$, and there is a minimum in the required field amplitude when $\omega_{\perp e} \sim 0$. In addition, the enduring mystery of non-zero penetration threshold at zero plasma natural frequency in EF experiments is resolved by two-fluid simulations. We find that the very small island and smooth bifurcation in EF penetration near zero frequency is hard to detect in the experiment, leading to a finite penetration threshold within the capability of the experimental measurements.

## 1. Introduction

Error fields (EFs), a kind of non-axisymmetric magnetic perturbations arising from imperfections in magnetic field-coils, have been a concern in magnetically confined fusion devices since the 1970's [1–3]. EFs with very small magnitude, typically $b_r/B_t \sim 10^{-4} - 10^{-3}$, can penetrate through resonant surfaces and drive magnetic reconnection in a tearing-stable plasma, generating locked magnetic island chains (named as "locked modes") [4]. Here $b_r$ is the resonant harmonic of the radial EF, and $B_t$ is the axisymmetric toroidal magnetic field. These locked modes severely degrade global energy confinement [5] and frequently cause major disruptions [1–3], leading to unacceptable limitations on the available operating space in tokamak experiments. To mitigate EF caused locked modes, EF correction (EFC) coil sets are commonly used to compensate the intrinsic EFs [3] and are also planned in ITER [3]. A particular concern for the EFC coils design in ITER is determining its expected tolerance to EFs (the EF penetration threshold), which has to be answered by the extrapolation from EF studies in recent tokamaks.

A series of experimental investigations have been performed on existing tokamaks [1, 2, 6–21] to obtain an empirical scaling of the EFs penetration threshold with basic plasma parameters. In such experiments, resonant magnetic perturbations (RMPs) generated by coils [2] are actively used to study the dependence of EF penetration threshold $b_r$ on plasma parameters. The scaling has been roughly derived in the form of $b_r/B_t \propto n_e^{\alpha_n} q_{95}^{\alpha_q} R_0^{\alpha_R}$ with $\alpha_n \sim 1$, $\alpha_B \sim -2.9$ to $-1$, $\alpha_q \sim 0$ to 1.6, $\alpha_R$ satisfies $8\alpha_n + 5\alpha_B - 4\alpha_R = 0$ [3]. Here, $n_e$ is the density, $q_{95}$ is the safety factor at the surface containing 95% of the poloidal magnetic flux, $R_0$ is the major radius, $\alpha_n$, $\alpha_B$, $\alpha_q$ and $\alpha_R$ are the scaling coefficients on these parameters. The tolerable EF level in ITER can be simply estimated based on this empirical scaling and the expected parameters in ITER. However, more experiments reveal additional uncertainties in this empirical scaling law:

(i) The scaling can be substantially different, i.e. $\alpha_n \sim 1$ is observed in DIII-D [1, 7], JET [7, 8], Alcator C-MOD [10],TEXTOR [11], MAST [14] and NSTX [16], while $\alpha_n \sim 0.5$ is also observed in COMPASS-C [2], JET [22], NSTX [17], J-TEXT [18] and EAST [21].

(ii) The separate single-variable scanning in the experiments is actually multi-variables scanning due to the coupling of plasma parameters [21], i.e. both electron temperature and plasma rotation vary during $n_e$ scanning, electron temperature also varies during $B_t$ scanning, etc.

(iii) Dependence on both the plasma rotation and electron temperature is missing in the empirical scaling as is its measurement. This is emphasized in theory as discussed in the following [4].

(iv) $b_r$ calculated from vacuum magnetic perturbations is used in the scaling in most of the experiments, while the plasma response is found to be important [23–25] and has not been considered except several recent studies [16, 17, 21, 26].

These limitations for the empirical scaling enlarge the uncertainty to predict EF tolerance in ITER.

In the meantime, the theoretical model of EF penetration has been developed to offer a coherent explanation for the experimental signatures of EFs penetration. The model based on magnetohydrodynamics (MHD) treatment (single-fluid) developed by Fitzpatrick [4] built the physics basis of EF penetration and predicted the scaling of EF penetration threshold on plasma parameters. The scaling of the EF penetration threshold predicted by this model takes the form [4]

$$b_r/B_t \propto n_e^{7/12}\, \tau_\mu^{-7/12} T_e^{5/8} \omega \;, \tag{1}$$

in the so-called *visco-resistive* regime, and the critical threshold in the *Rutherford* regime is [4]

$$b_r/B_t \propto n_e^{3/5}\, \tau_\mu^{-3/5} T_e^{3/5} \omega \,, \tag{2}$$

and the scaling differs in several other regimes [4] we do not show here. Here, the viscous time $\tau_\mu = a^2/\mu$, and $\mu$ is the plasma viscosity. The scaling of EF threshold in equations (1) and (2) show the plasma parameters that determines EF penetration threshold. The EF scaling law from this reduced MHD model agrees well with experimental observations in Ohmic or L-mode plasma [2]. However, the scaling shows no density dependence when the viscosity diffusion time $\tau_\mu$ adopting the Neo-Alcator scaling $\tau_\mu \propto \tau_E \propto n_e$ [27] as mentioned in [28] and is, therefore, inconsistent with experiments. Linear driftMHD model (two-fluid) [29, 30] of EF penetration is developed to take into account the effect of high temperature, however, the scaling dependence on density is still too weak ($\alpha_n \sim 0.25 - 0.5$) to account for the experimental observations. A later developed nonlinear drift-MHD model [28] predicts a linear dependence of the EF threshold with density, which is consistent with those observations with $\alpha_n = 1$. Nonlinear two-fluid modeling further reveals that the EF threshold significantly increases as the perpendicular flow frequency increases [31]. It seems that each theory model works well predicting the exponent on some scaling parameters, but poorly for others. To date, no MHD modeling has reproduced or validated this theoretical scaling. In addition to fusion plasma, scaling and similarity technique is also a useful tool for developing and testing reduced models of complex phenomena, including the dynamics of plasma phenomena [32–35].

The objective of this paper is to examine the difference in EF scaling between *single-fluid* and *two-fluid* models by using nonlinear MHD modeling, and to provide clarification of theoretical expectations for ongoing experiments as well as extrapolations to ITER. Compared to those theoretical studies [4, 28–30], it is the first time that a numerical MHD code simulates the scaling law of EF penetration threshold. The simulated EF scaling law is very close to the analytical scaling law in both the *visco-resistive* and *Rutherford* regimes [4]. The good consistency between single-fluid scaling and analytical theory validates the capability of the TM1 code simulating EF scaling. The scaling laws from the two-fluid simulations reveal how they differ from single-fluid scalings depending on the plasma parameters. Both the single-fluid and twofluid simulations highlight the importance of electron temperature in EF scaling. In addition, an enduring mystery in EF experiments is that the penetration threshold is non-zero even when the plasma natural frequency is zero. We resolve this mystery by two-fluid simulations that the very small island and smooth bifurcation in EF penetration near zero frequency is hard to detect in the experiment, leading to a finite penetration threshold within the capability of the experimental measurements.

The paper is structured as follows. First, in section 2, the TM1 numerical model is introduced. The EF scaling is modeled by scanning the plasma parameters ($n_e, T_e, B_t, \tau_\mu, R_0$ and rotation) in a wide region covering recent devices as well as ITER as presented in section 3. Single-fluid modeling of EF scaling in section 3.1 is found to be similar to that of the single-fluid analytical model [4]. Then EF scaling is modeled by using nonlinear two-fluid simulations in section 3.2, and the modeled scaling coefficients $\alpha_n, \alpha_T$, $\alpha_\mu$ and $\alpha_B$ change significantly as the plasma rotation frequency $\omega_E$ changes. Finally, the dependence of EF threshold on perpendicular rotation frequency $\omega_{\perp e}$ is modeled, which shows a linear dependence and the EF threshold is near zero when $\omega_{\perp e} \sim 0$. The concluding discussion, brief prediction for ITER and summary are given in section 4.

## 2. Numerical model

The TM1 model uses a straight cylindrical, circular cross section tokamak for simplicity when simulating the nonlinear evolution of EF penetration. The magnetic field is defined as $B = B_t e_t + \nabla\psi \times e_t$, where $\psi$ is the magnetic flux function. The plasma velocity is defined as $v = v_{||} e_{||} + \nabla\varphi \times e_t$, where $v_{||}$ and $v_\perp = \nabla\varphi \times e_t$ are the parallel and perpendicular velocity, respectively. The cold ion assumption is made as in [36]. The basic equations utilized here are Ohm's law, the equation of motion in the parallel and the perpendicular direction (after taking the operator $e_t \cdot \nabla \times$), the electron continuity equation and the energy conservation equation [37]. Normalizing length to the minor radius $a$, the time $t$ to the resistive time $\tau_R = a^2\mu_0/\eta$, the helical flux $\psi$ to $aB_t$, the velocity $v$ to $a/\tau_R$, the density $n_e$ and the electron temperature $T_e$ to their values at the EF resonant surface, these equations become [38, 39]

$$\frac{d\psi}{dt} = E - \eta j + \Omega(\nabla_{||} n_e + \nabla_{||} T_e), \quad (3)$$

$$\frac{dU}{dt} = -S^2\nabla_{||}j + \mu\nabla^2_\perp U + S_m, \quad (4)$$

$$\frac{dv_{||}}{dt} = -c_s^2\nabla_{||}P/n_e + \mu\nabla^2_\perp v_{||}, \quad (5)$$

$$\frac{dn_e}{dt} = d_1\nabla_{||}j - \nabla_{||}(n_e v_{||}) + \nabla\cdot(D_\perp\nabla n_e) + S_n, \quad (6)$$

$$\frac{3}{2}n_e\frac{dT_e}{dt} = d_1 T_e\nabla_{||}j - T_e n_e\nabla_{||}v_{||} + n_e\nabla\cdot(\chi_{||}\nabla_{||}T_e) + n_e\nabla\cdot(\chi_\perp\nabla_\perp T_e) + S_p, \quad (7)$$

where $d/dt = \partial/\partial t + v_\perp\cdot\nabla$, $\nabla_{||}f = (B/B)\cdot\nabla f \approx (B/B_t)\cdot\nabla f$, $\nabla_\perp f = f^0 + (\partial f/\partial\theta)/r$, $\nabla^2{}_\perp f = (rf^0)^0 r - (\partial^2 f/\partial\theta^2)/r^2$ and the prime is $\partial/\partial r$. Plasma current density $j$ is derived according to $j = \nabla\times B$, $\eta$ is the normalized plasma resistivity and $E$ is the equilibrium electric field for maintaining the equilibrium plasma current density. $\Omega = \beta d_1$ determining the diamagnetic drift frequency, $\beta = 4n_eT_e/B^2$, $d_1 = \omega_{ce}/\mu_e$, $\omega_{ce}$ and $\mu_e$ are the electron cyclotron and collisional frequency, respectively. The magnetic Reynolds number

$S = \tau_R/\tau_A$, where $\tau_A = a/V_A$ is the toroidal Alf`en time. $U = -\nabla^2{}_\perp\varphi$ is the plasma vorticity, $\mu$ is the plasma viscosity, $c_s = (T_e/m_i)^{1/2}$ being the ion sound velocity and $P$ the plasma pressure. $\chi_{||}$ and $\chi_\perp$ are the parallel and perpendicular heat conductivities, $D_\perp$ is the perpendicular particle transport coefficient. $S_m$, $S_n$ and $S_p$ are the source of momentum, particle and heating power, which lead to an equilibrium profile of plasma rotation $\omega$, density $n_e$ and temperature $T_e$.

Equations (3)-(7) are the coupled two-fluid equations for modelling the nonlinear evolution of the drifting tearing mode [38, 40], which reduce to the single-fluid MHD equations if $\Omega = 0$ is taken in Ohm's law to neglect the parallel electron temperature and the density gradient. The two-dimensional electron heat transport is included self-consistently in equations (3)–(7). Equations (3)–(7) are solved simultaneously based on implicit differential method using the initial value code TM1, which has been used for modelling the stability of the drifting tearing mode earlier [38, 40], the nonlinear growth and saturation of the neoclassical tearing mode (NTM) and their stabilization by electron cyclotron current drive (ECCD) [41], resonant field penetration, particle and energy transport in the TEXTOR [31, 39], DIII-D [42, 43] and J-TEXT [44, 45] tokamaks. Dedicated numerical methods are utilized in the code to keep the numerical error at a very low level even for high values of $S$ and $\chi_{||}/\chi_\perp$ [46].

The calculations in this work only include the $m/n$ = 3/2 resonant helicity perturbations (non-resonant components are not included). In addition to the fundamental harmonic, higher harmonic perturbations (including from $2^{nd}$ to $7^{th}$ harmonics) as well as the change in the equilibrium quantities (the $m/n$ = 0/0 component) are self-consistently calculated. The toroidal magnetic field $B_t$ is taken to be a constant and the toroidal mode coupling is neglected. Fourier decomposition in the poloidal and toroidal directions and finite differences along the radial direction are utilized in the code. The calculation region is from the magnetic axis at $\psi_N = 0$ to the plasma edge at $\psi_N = 1$. The boundary conditions are as the following [31].

(i) The radial gradients of all quantities are zero at $\psi_N = 0$.

(ii) All the perturbations ($m/n$ 6= 0/0) are zero at $\psi_N = 1$ except for the $m/n$ = 3/2 magnetic perturbation given by the following equation (8) to take into account the EF.

(iii) All the equilibrium ($m/n$ = 0/0) quantities take the same value as the original equilibrium ones at $\psi_N = 1$.

The effect of the EF is considered by the boundary condition

$$\psi_{m/n}(\psi_N = 1) = \psi_a a B_t \cos(m\theta + n\phi + \Phi_0), \quad (8)$$

where $\psi_a$ and $\Phi_0$ describe the amplitude and phase of the applied EF of the $m/n$ component at $\psi_N = 1$. The radial EF at $\psi_N = 1$ is given by $b_r = m\psi_a B_t$. The driving EF is taken from and kept at the RMP value at the plasma boundary for each harmonic

## 3. Numerical results

TM1 simulations of $m/n$=3/2 EF penetration are performed by using the equilibrium and profiles from the experimental DIII-D L-mode discharge 171672. The input parameters are based on the experimental parameters and equilibrium profiles of shot 171672 as shown in figure 1. The toroidal magnetic field on axis is -1 T and the plasma minor and major radii are $a$ = 0.63 m and $R$ = 1.7 m. The $q$ = 3/2 rational surface is located at $\psi_N$ = 0.6, and $q_{95}$ = 3.3. The transport coefficients are obtained from TRANSP [47] power and particle balance calculations based on the profile measurements, which are $\mu \sim \chi_\perp \sim 2D_\perp \sim 0.5\ m^2/s$. To model EF scaling on plasma parameters, the electron density $n_e$ at $q$ = 3/2 surface is scanned from $0.5\times10^{19}m^{-3}$ to $5.1\times10^{19}m^{-3}$, the electron temperature at $q$ = 3/2 surface is scanned from 50 eV to 1200 eV, the toroidal field $B_t$ is scanned from 1 T to 5.5 T, major radius $R_0$ is scanned from 1 m to 6.5 m, and the plasma viscosity $\mu$ is scanned from 0.05 $m^2/s$ to 5 $m^2/s$. Here, each parameter with more than 20 points are scanned. These parameters lead to a wide range of the input parameters in equations (3)-(7), for example, the magnetic Reynolds number S ranges from $4 \times 10^5$ to $5 \times 10^7$. In the following, the scaling of EF threshold on plasma parameters are modeled by using *single-fluid* and *two-fluid* simulations in section 3.1 and 3.2.

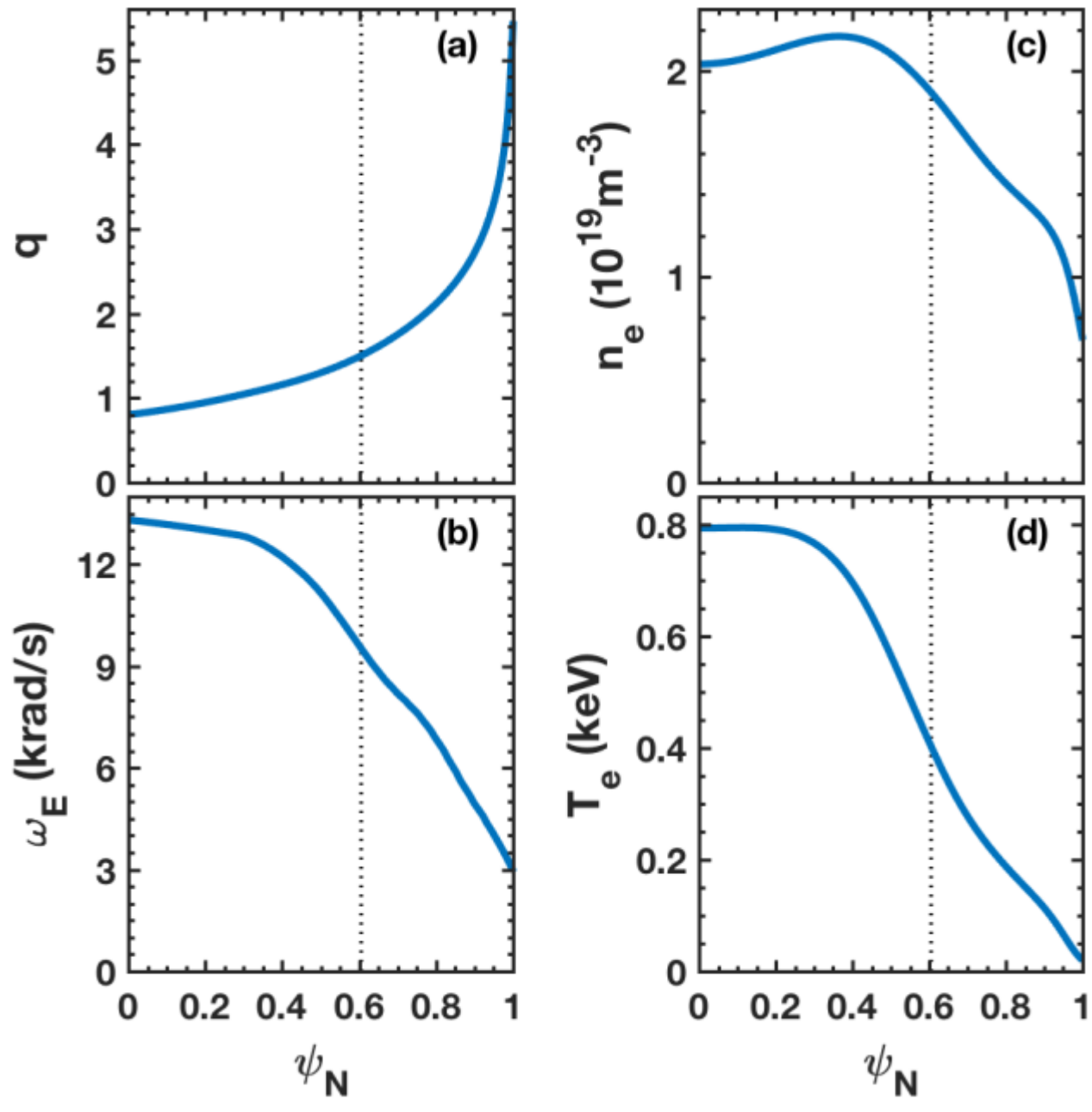

**Figure 1.** Equilibrium profiles of (a) safety factor $q$, (b) $E\times B$ rotation frequency $\omega_E$, (c) electron density $n_e$ and (d) temperature $T_e$ from DIII-D L-mode discharge #171672 are used for modeling. The location of $q$ = 3/2 rational surface is indicated by the dotted lines.

## *3.1.* Single-fluid scaling of EF threshold

The coupled two-fluid equations (3)-(7) reduce to the single-fluid MHD equations if $\Omega$ = 0 is taken in Ohm's law to neglect diamagnetic drift effect. In the single fluid model, it decouples both the particle and energy transport, and the bifurcation of plasma response to EF from screening to penetration is governed by the Ohm's law in equation (3) and motion equation (4). It is clear that the parameters including $\eta$ ($\eta \propto T_e^{-3/2}$), $n_e$, $\omega$, $\mu$, $B_t$ and $a$ are involved in equations (3) and (4) and hence possibly determine the scaling of EF penetration threshold. In this section, results of EF scaling from single-fluid simulation are presented.

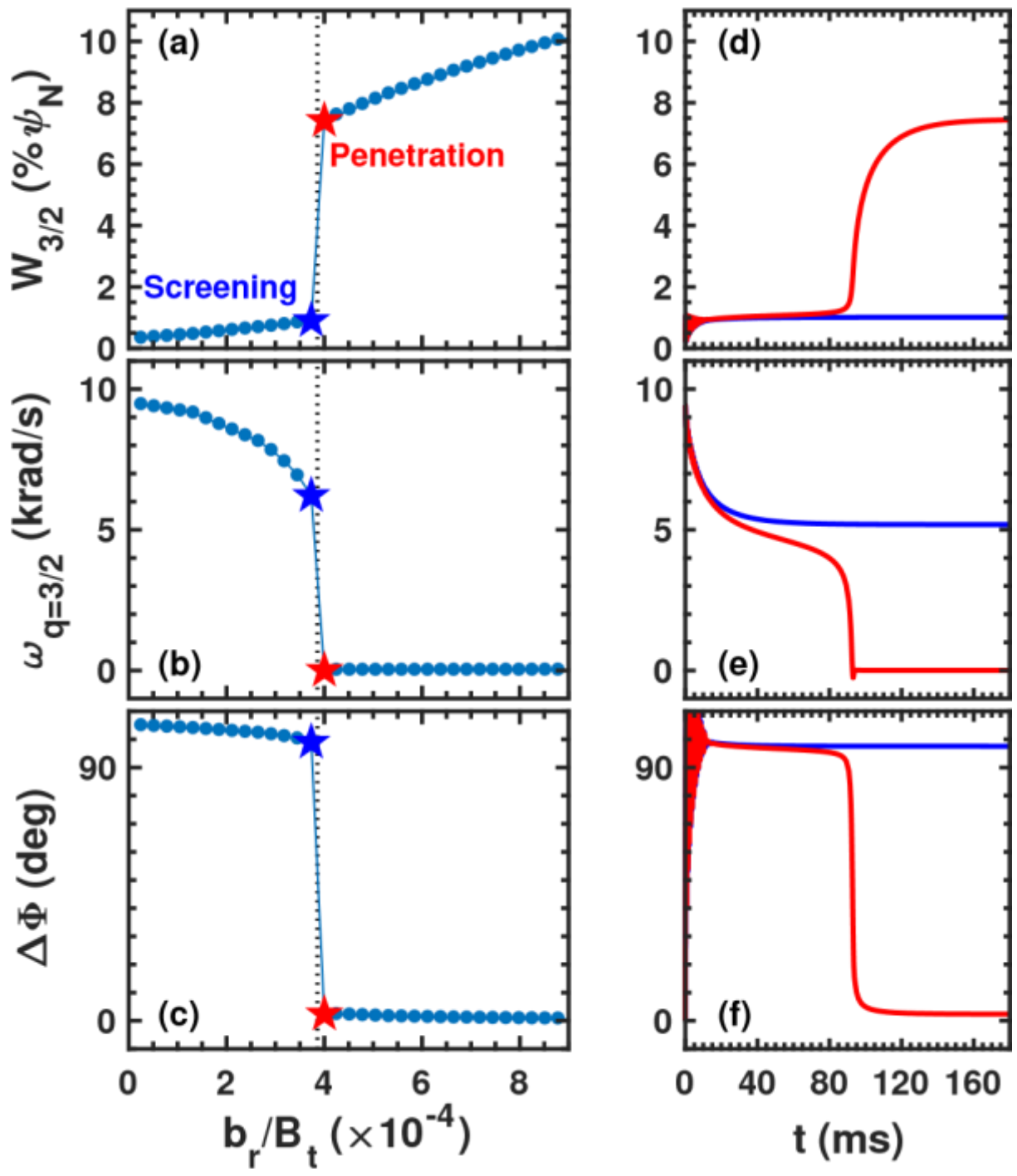

**Figure 2.** $m/n$ = 3/2 EF penetration case. TM1 simulated (a) island width $W_{3/2}$, (b) rotation frequency $\omega_{q=3/2}$ and (c) phase difference $\Delta\Phi = \Phi - \Phi_0$ between the plasma response field ($\Phi$) and vacuum field ($\Phi_0$) versus normalized radial magnetic perturbation $b_r/B_t$. Time evolution of (d) $W_{3/2}$, (e) $\omega_{q=3/2}$ and (f) $\Delta\Phi$ around the bifurcation from screening (blue) to penetration (red), the RMP strength corresponds to blue and red stars in (a)-(c). Here, the simulation starts at $t$ = 0 with constant EF.

In the experiment, the RMPs strength is usually ramping up slowly to trigger EF penetration and the $b_r$ amplitude at the onset of locked modes is considered as the penetration threshold. The finite ramp rate of $b_r$ usually causes an error in the EF threshold. In our simulations, constant EF is applied and its amplitude is scanned separately, a typical example of such scanning is shown in figure 2. In this case, $n_e = 1.9\times10^{19}\ m^{-3}$ and $T_e = 0.4\ keV$ at the $q$ = 3/2 surface are used, which leads to magnetic Reynolds number $S = 1.2\times10^7$. The saturated island width $W_{3/2}$, the rotation frequency $\omega_E$ at $q$ = 3/2 surface and the phase difference $\Delta\Phi$ between the magnetic island and EF field are shown as a function of EF amplitude in figure 2(a)-2(c). Very weak EF is shielded by the plasma as indicated by a small, saturated island width ($< 0.01\psi_N$) and $\Delta\Phi > 90^\circ$. The screening current at the resonant surface drives resistive kink response and generates nonzero flux perturbation, from which a small island width is derived [44]. As the EF amplitude increases, the increasing electromagnetic (EM) force decreases the plasma rotation and $\Delta\Phi$ as shown in figure 2(b) and 2(c). When EF amplitude exceeds the threshold $b_r/B_t = 4\times10^{-4}$, bifurcation from screening to penetration happens with a fast jump in the saturated island width (figure 2(a)), and both the rotation at the $q$ = 3/2 surface and $\Delta\Phi$ decreasing to near zero. The detailed evolution of $W_{3/2}$, $\omega_{q=3/2}$ and $\Delta\Phi$ for EF just below (blue star) and above (red star) the penetration threshold are shown in blue and red curves in figure 2(d)-2(f), respectively. The red curves in figure 2(d)-2(f) show a very fast process ($\sim 10\ ms$) of bifurcation from screening to penetration with fast growth in the island width $W_{3/2}$ and fast slowing down of plasma rotation $\omega_{q=3/2}$, which are consistent with experimental observations in DIII-D [20].

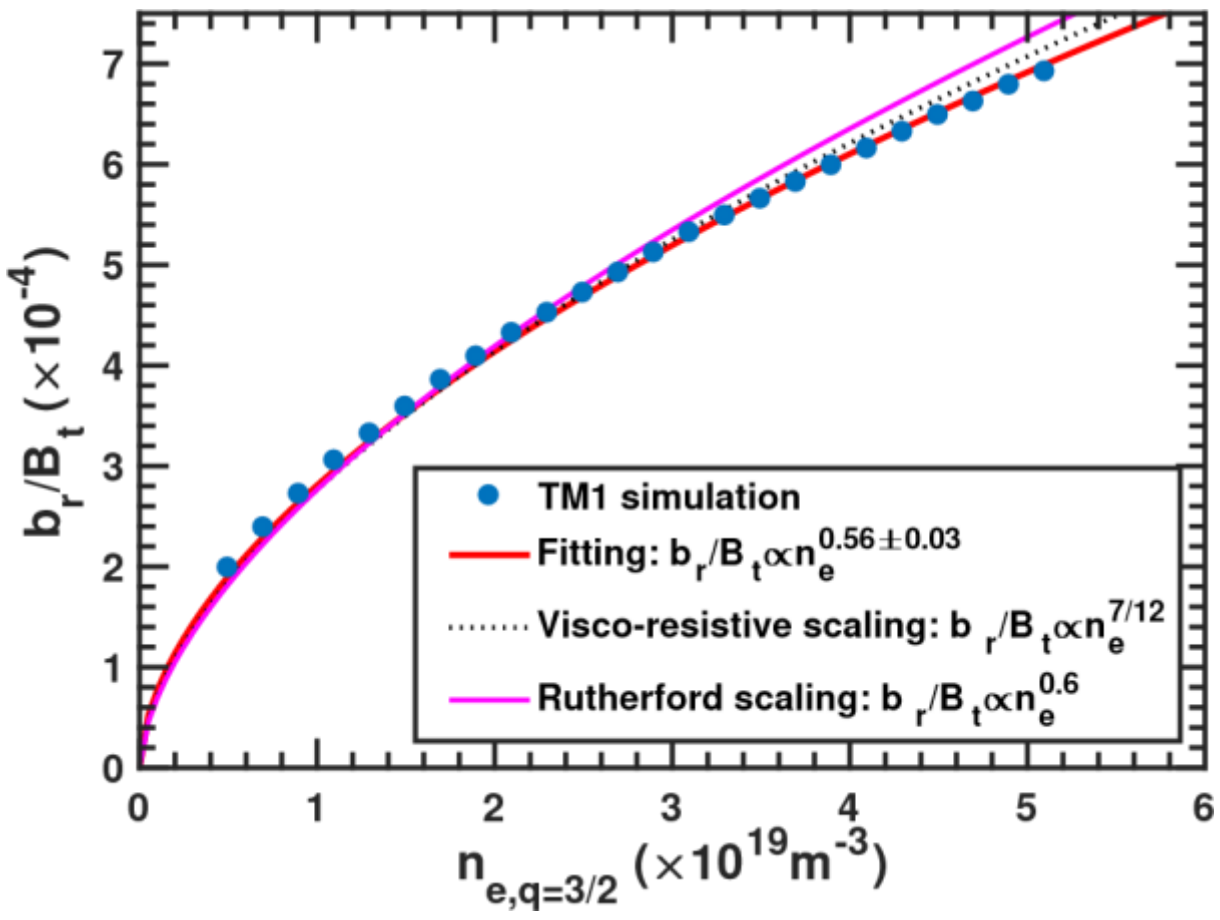

**Figure 3.** TM1 single-fluid simulation of $m/n$ = 3/2 EF penetration threshold versus electron density at $q$ = 3/2 surface $n_{e,q=3/2}$. A least-squares fitting (red curve) for the numerical results indicates a density scaling of $b_r/B_t \propto n^0{}_e{}^{.56\pm0.03}$. The analytical scaling on density from the visco-resistive regime (Equation (1), black dotted curve) and Rutherford regime (Equation (2), purple solid curve) are shown for comparison.

The density profile in figure 1(c) is proportionally changed to scan the dependence of EF threshold on $n_{e,q=3/2}$ ranging from 0.5 × 10$^{19}$ $m^{-3}$ to 5.1 × 10$^{19}$ $m^{-3}$ as shown in figure 3. In this scanning, other parameters including $\omega_{E,q=3/2}$ = 10 krad/s, $B_t$ = 1 T, $T_{e,q=3/2}$ = 0.4 keV, and $\mu = \chi_\perp = 2D_\perp = 0.5$ $m^2/s$ are kept unchanged to make sure only $n_e$ is involved. Figure 3 shows that the nonlinear TM1 modeled EF threshold (blue circles) increases with the increasing density, and a least-squares fitting is applied as shown by the red curve, which indicates $b_r/B_t \propto n_e^{0.56\pm0.03}$. Figure 3 reveals that for higher density, the required EF penetration threshold will be stronger, the reason is that higher density leads to higher plasma inertia and hence it requires stronger EF to slow down the plasma rotation to trigger field penetration. The scaling shown in figure 3 is very close to experimental observations in Ohmic [2, 18] or L-mode [17, 21, 22] discharges. Besides, the analytical scaling on density in both the visco-resistive and Rutherford regimes (Equations (1) and (2)) is also shown in figure 3, it is clear that the simulated density scaling is very close to the theoretical scaling.

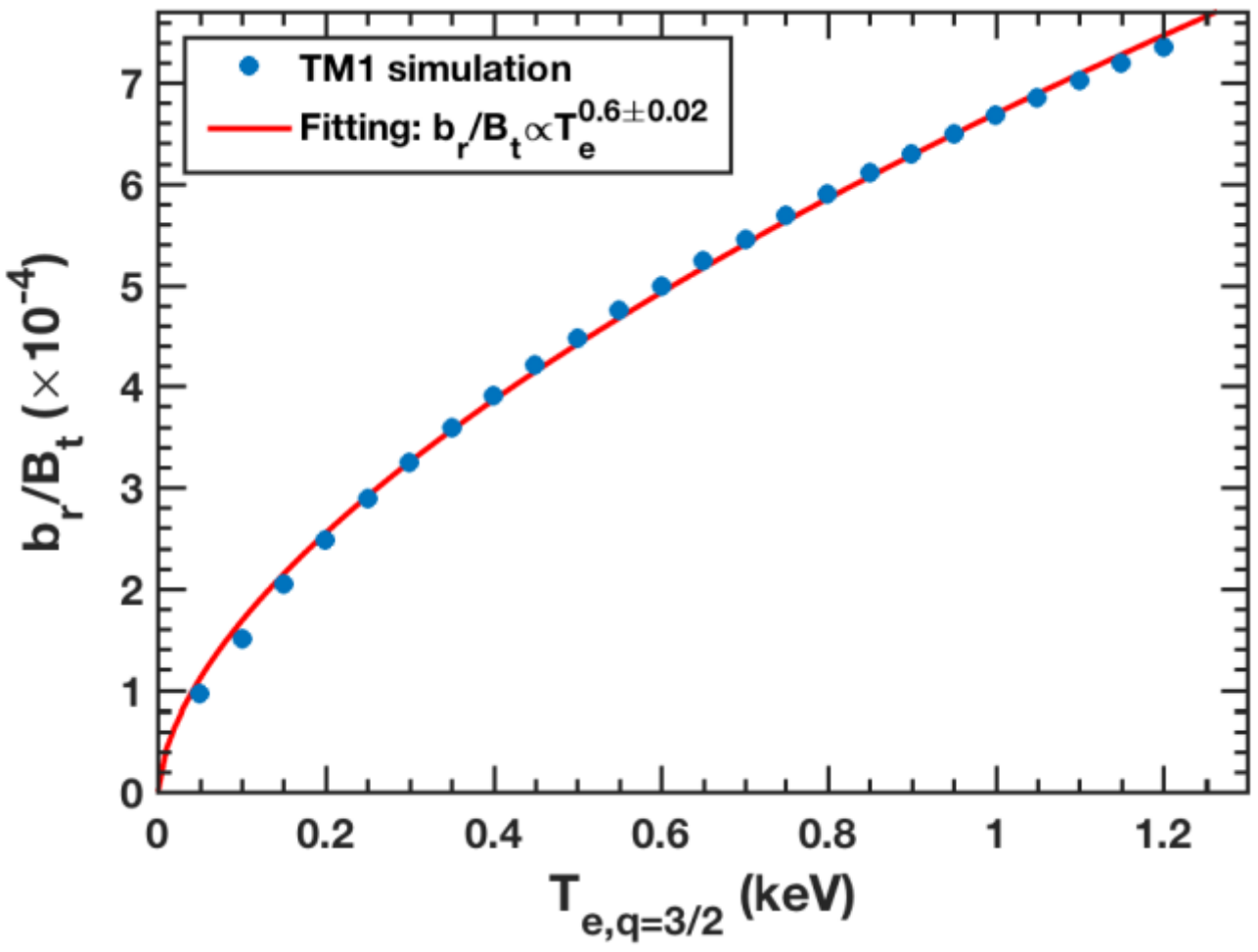

**Figure 4.** TM1 single-fluid simulation of $m/n$ = 3/2 EF penetration threshold versus temperature at $q$ = 3/2 surface $T_{e,q=3/2}$. A least-squares fitting (red curve) for the numerical results indicates a temperature scaling of $b_r/B_t \propto T_e^{0.6\pm0.02}$. The analytical scaling on temperature from the visco-resistive regime (Equation (1), black dotted curve) and Rutherford regime (Equation (2), purple solid curve) are shown for comparison.

The temperature profile in figure 1(d) is proportionally changed to scan the dependence of EF threshold on $T_{e,q=3/2}$ ranging from 50 $eV$ to 1200 $eV$ as shown in figure 4, while the parameters (other parameters including $\omega_{E,q=3/2}$ = 10 krad/s, $B_t$ = 1 T, $n_{e,q=3/2}$ = 1.9 × $10^{19}$ $m^{-3}$, and $\mu = \chi_\perp = 2D_\perp$ = 0.5 $m^2/s$) are also kept unchanged. Figure 4 shows that the nonlinear TM1 modeled EF threshold (blue circles) increases with the increasing $T_{e,q=3/2}$, and a least-squares fitting is applied as shown by the red curve, which indicates $b_r/B_t \propto T_e^{0.6\pm0.02}$. Figure 4 reveals that for higher temperature, the required EF penetration threshold will be stronger, the reason is that higher temperature leads to lower plasma resistivity $(\eta \propto T_e^{-3/2})$, which gives rise to stronger screening on EF and hence requires stronger EF to slow down the plasma rotation to trigger field penetration. Interestingly, the simulated scaling on temperature is very close to the analytical scaling on temperature in both the visco-resistive and Rutherford regimes (Equations (1) and (2)) as shown in figure 4. It should be noted that EF scaling on temperature is missing in most of the experimental scaling, while the strong positive dependence in figure 4 indicates that the temperature effect should be taken into account for EF scaling.

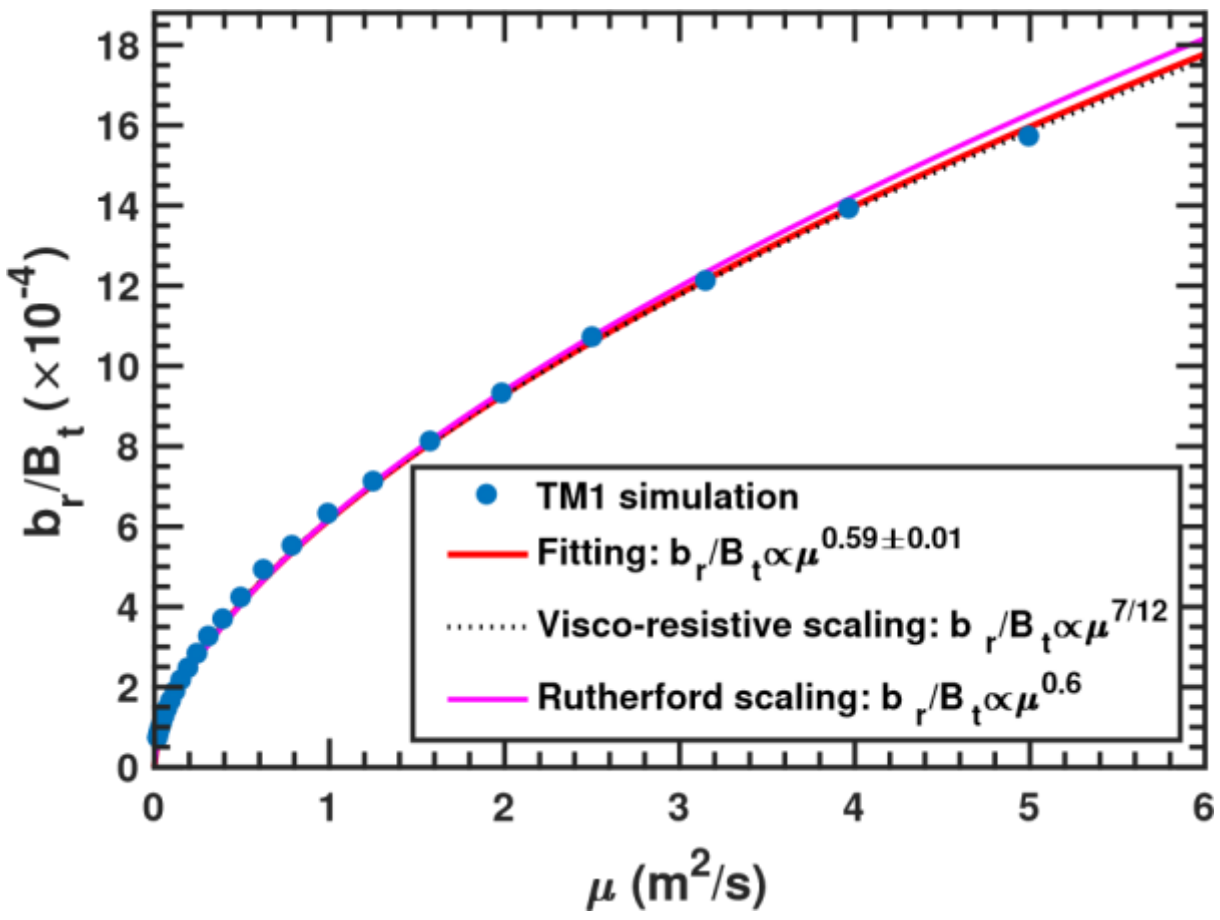

**Figure 5.** TM1 single-fluid simulation of $m/n$ = 3/2 EF penetration threshold versus plasma viscosity $\mu$. A least-squares fitting (red curve) for the numerical results indicates a viscosity scaling of $b_r/B_t \propto \mu^{0.59\pm0.01}$. The analytical scaling on viscosity from the visco-resistive regime (Equation (1), black dotted curve) and Rutherford regime (Equation (2), purple solid curve) are shown for comparison.

The dependence of EF threshold on plasma viscosity is modeled by scanning $\mu$ from 0.05 $m^2/s$ to 5 $m^2/s$ as shown in figure 5, while the parameters (other parameters including $\omega_{E,q=3/2}$ = 10 krad/s, $B_t$ = 1 T, $n_{e,q=3/2}$ = 1.9 × $10^{19}$ $m^{-3}$, $T_e$ = 0.4 keV and $\chi_\perp = 2D_\perp$ = 0.5 $m^2/s$) are kept unchanged. Figure 5 shows that the nonlinear TM1 modeled EF threshold (blue circles) increases with the increasing $\mu$, and a least-squares fitting is applied as shown by the red curve, which indicates $b_r/B_t \propto \mu^{0.59\pm0.01}$. Since the viscous time scale $\tau_\mu = a^2/\mu$, the scaling can be expressed as $b_r/B_t \propto \tau_\mu^{-0.59\pm0.01}$. Figure 5 reveals that higher viscosity leads to higher EF penetration threshold, the reason is that higher viscosity leads to stronger damping on the EM torque and hence requires stronger EF to slow down the plasma rotation to trigger field penetration. Again, the simulated scaling on viscosity is very close to the analytical scaling in both the visco-resistive and Rutherford regimes (Equations (1) and (2)) as shown in figure 5.

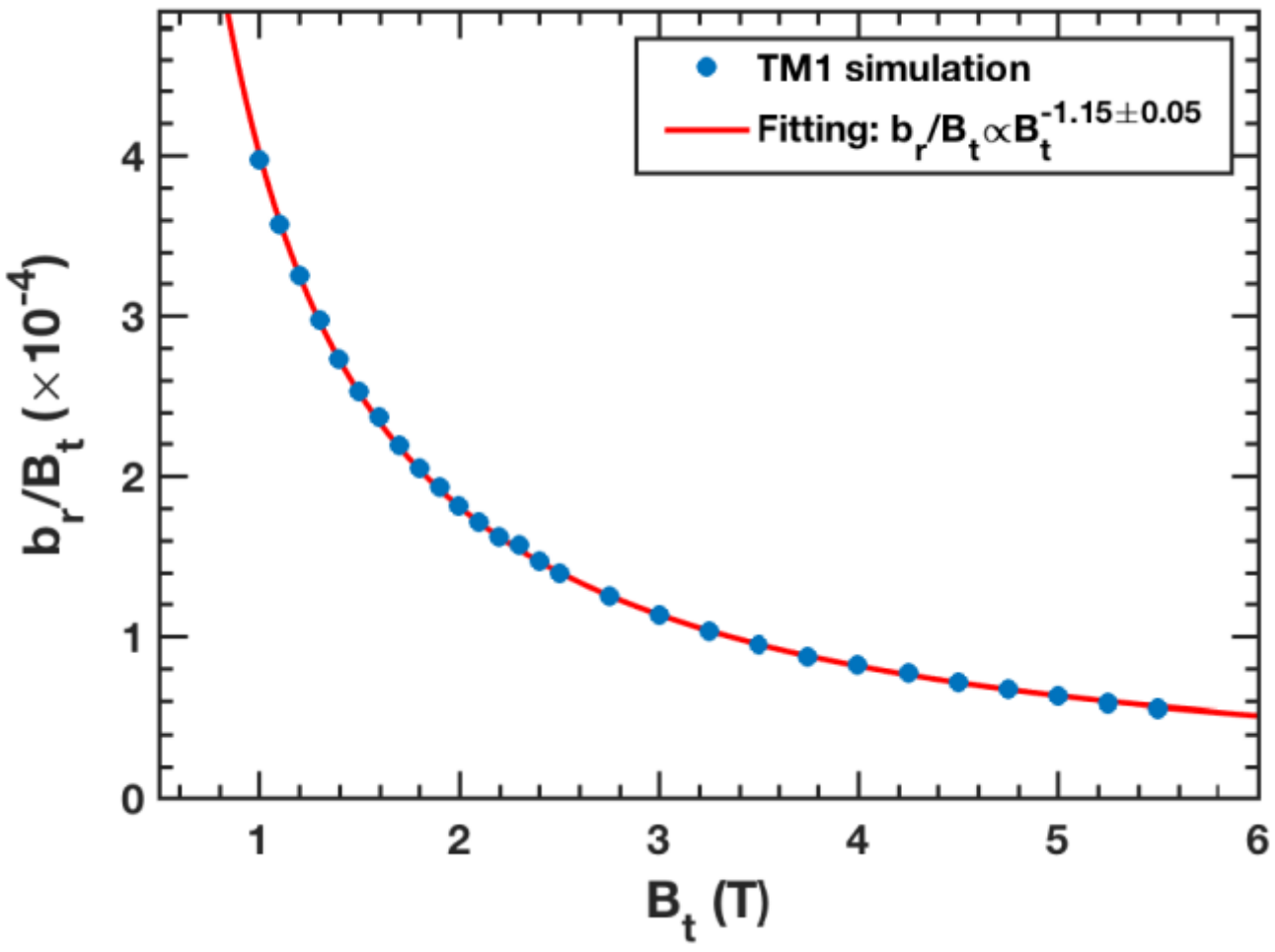

**Figure 6.** TM1 single-fluid simulation of $m/n$ = 3/2 EF penetration threshold versus toroidal field $B_t$. A least-squares fitting (red curve) for the numerical results indicates a toroidal field scaling of $b_r/B_t \propto B_t^{-1.15\pm0.05}$.

The dependence of EF threshold on toroidal field is modeled by scanning $B_t$ from 1 T to 5.5 T as shown in figure 6, while the parameters ($\omega_E$, $n_e$, $T_e$ and $\mu$) are kept unchanged. This scanning covers $B_t$ from typical current devices (∼ 1−3 T) to ITER (5.2 T). Figure 6 shows that the TM1 modeled normalized EF threshold $b_r/B_t$ (blue circles) decreases with the increasing $B_t$, and a least-squares fitting is applied as shown by the red curve, which indicates $b_r/B_t \propto B_t^{-1.15\pm0.05}$. This scaling in figure 6 is very close to the observations in DIII-D [9] and JET [6, 7] but substantially differs to observation in COMPASS-C ($\alpha_B = -2.9$) [2]. In addition, figure 6 shows that the normalized EF threshold decays very quickly when $B_t$ increases from 1 T to 3 T, revealing that there will be a significant uncertainty in toroidal field scaling owing in a typical experimentally limited range of $B_t$ variation. Therefore, cross-machine scaling with a range of $B_t$ is necessary to obtain an accurate $\alpha_B$.

The dependence of EF threshold on major radius is modeled by scanning $R_0$ from 1 m to 6.5 m as shown in figure 7, the aspect ratio $a/R_0$ is kept the same as that of DIII-D, and other parameters ($\omega_E$, $n_e$, $T_e$, $B_t$ and $\mu$) are also kept unchanged. This scanning covers $R_0$ from recent devices (∼ 1 − 3 m) to ITER (6.5 m). The variation of machine size affects the time scales determining the dynamics of EF penetration, i.e. $\tau_R$, $\tau_\mu$ and $\tau_A$. Figure 7 shows that, for plasma with constant angular rotation frequency $\omega$, the nonlinear TM1 modeled EF threshold (blue circles) increases with the increasing $R_0$, and a least-squares fitting indicates $b_r/B_t \propto R_0^{0.78\pm0.03}$. However, for constant velocity $V$, the increasing major radius leads to a smaller angular rotation frequency $\omega$ and hence leads to a smaller EF threshold to penetration (red circles), which scales as $b_r/B_t \propto R_0^{-0.18\pm0.03}$. Theoretical prediction shows that the intrinsic rotation may probably decrease when the device size increases [48, 49], indicating the plasma velocity may even decrease with increasing $R_0$. As a result, $\alpha_R$ may be more negative than $\alpha_R = -0.18$.

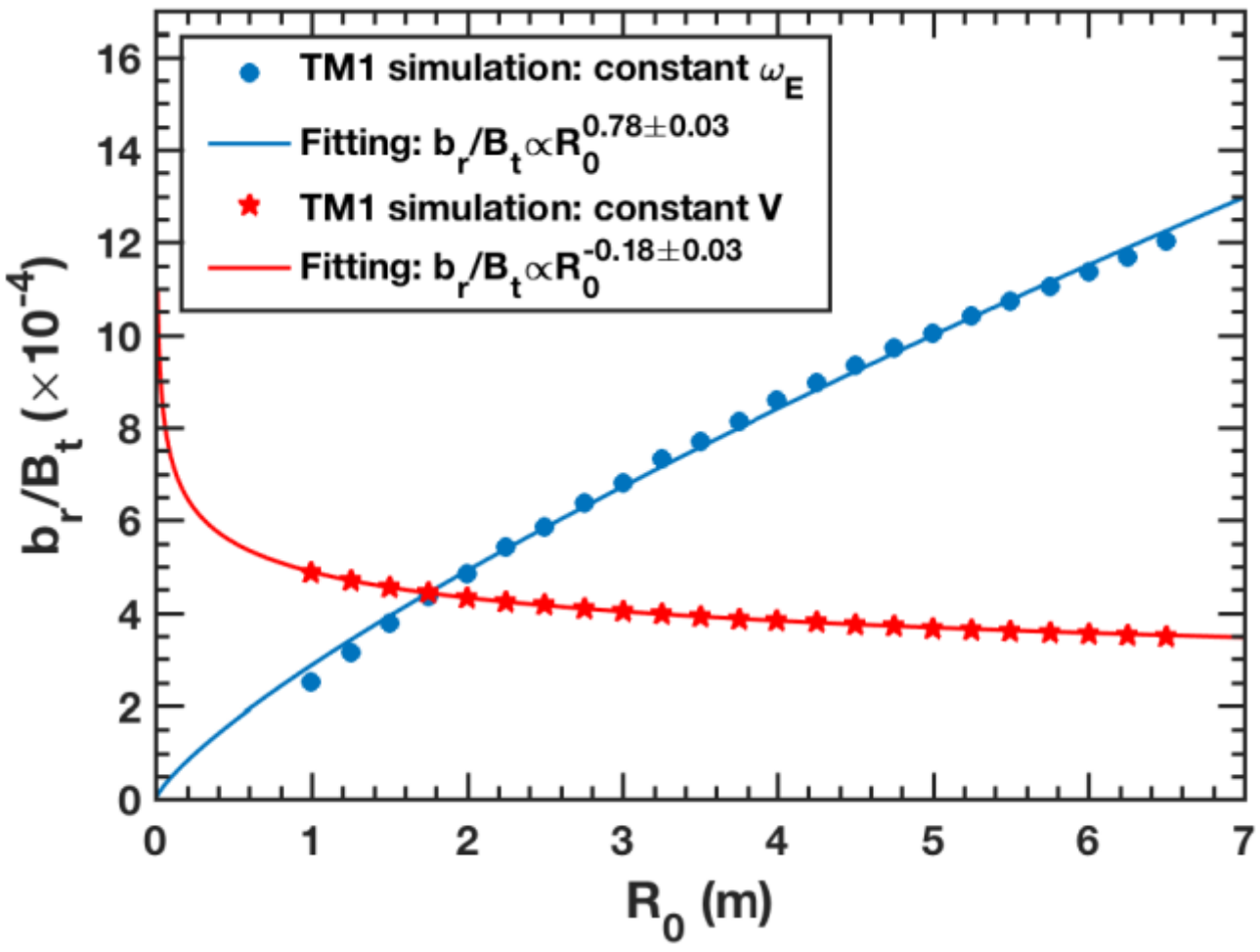

**Figure 7.** TM1 single-fluid simulation of $m/n = 3/2$ EF penetration threshold versus major radius $R_0$ for constant angular frequency $\omega_E$ (blue) and constant rotation velocity $V$ (red) at the $q = 3/2$ surface. Least squares fitting for the numerical results indicates a major radius scaling of $b_r/B_t \propto R_0^{0.78\pm0.03}$ for constant $\omega_E$ and $b_r/B_t \propto R_0^{-0.18\pm0.03}$ for constant $V$.

In the next section, the modeled scaling shows a linear dependence of the EF threshold on the natural rotation frequency $\omega$ at the resonant surface, which leads to a modeled scaling law together with the results shown in figures 3 to 6 following

$$b_r/B_t \propto n_e^{0.56\pm0.03}\,\tau_\mu^{-0.59\pm0.01} T_e^{0.6\pm0.02} B_t^{-1.15\pm0.05}\,\omega. \quad (9)$$

This modeled EF scaling is compared with reduced MHD theory as shown in table 1. We noted that the EF scaling on desnity $n_e$, temperature $T_e$, plasma viscous time $\tau_\mu$ and plasma rotation frequency $\omega$ in equation (9) is very close to the single-fluid theoretical scaling in both the *Rutherford* and *visco-resistive* regimes in equations (1) and (2). This EF scaling is also compared with DIII-D experiments with good consistency [26]. To our knowledge, it is the first time that nonlinear MHD modeling has reproduced the theoretical scaling of the EF threshold, which is a good validation for both the model and theory. However, the modeled scaling also shows no density dependence when $\tau_\mu \propto \tau_E \propto n_e$ according to the Neo-Alcator scaling [27] in the linear Ohmic confinement regime and is, therefore, inconsistent with experiments, indicating that single-fluid scaling is insufficient to address the EF scaling. In addition, taking into account the high temperature effect in the tokamak plasma, it is necessary to extend the single-fluid to two-fluid modeling based on the good agreement between TM1 simulation and theoretical scaling.

**Table 1:** Comparison of the EF scaling between reduced MHD theory [4] and TM1 single-fluid modeling.

| $\alpha$ | *Rutherford regime* | *Visco-resistive regime* | TM1 |
|---|---|---|---|
| $\alpha_n$ | 0.6 | 0.583 | $0.56 \pm 0.03$ |
| $\alpha_T$ | 0.6 | 0.625 | $0.6 \pm 0.02$ |
| $\alpha_\mu$ | -0.6 | -0.583 | $-0.59 \pm 0.01$ |
| $\alpha_\omega$ | 1 | 1 | 1 |

## *3.2.* Two-fluid scaling of EF threshold

In this section, two-fluid modeling of the EF scaling is studied to identify how two-fluid effects affect the scaling. The coupled two-fluid equations (3)-(7) show that the plasma response to EFs couples the particle and energy transport according to the diamagnetic drift effect in the Ohm's law in equation (3), indicating the EF threshold not only depends on those parameters scanned in section 3.1 but also depends on those parameters involved in equations (6) and (7), i.e. $D_\perp$, $\chi_{||}$ and $\chi_\perp$. In the experiments, these transport coefficients are usually at the anomalous transport level, and their effects on EF scaling are briefly discussed in this section. In addition, in the two-fluid model, the natural frequency (perpendicular rotation frequency) consists of diamagnetic drift frequency $\omega_{*e} = (\partial P_e/\partial\psi)/n_e$ and the plasma rotation frequency $\omega_E = E_r/|RB_\theta|$ by $\omega_{\perp e} = \omega_{*e} + \omega_E$. The effect of EF on both the density and temperature profile will affect $\omega_{*e}$, as a result, the EF scaling will be sensitive to the relative magnitude and direction of $\omega_E$ to $\omega_{*e}$. Consequently, the EF scaling on $n_e$, $T_e$ and $\mu$ will be studied with different ratios of $\omega_E/\omega_{*e}$.

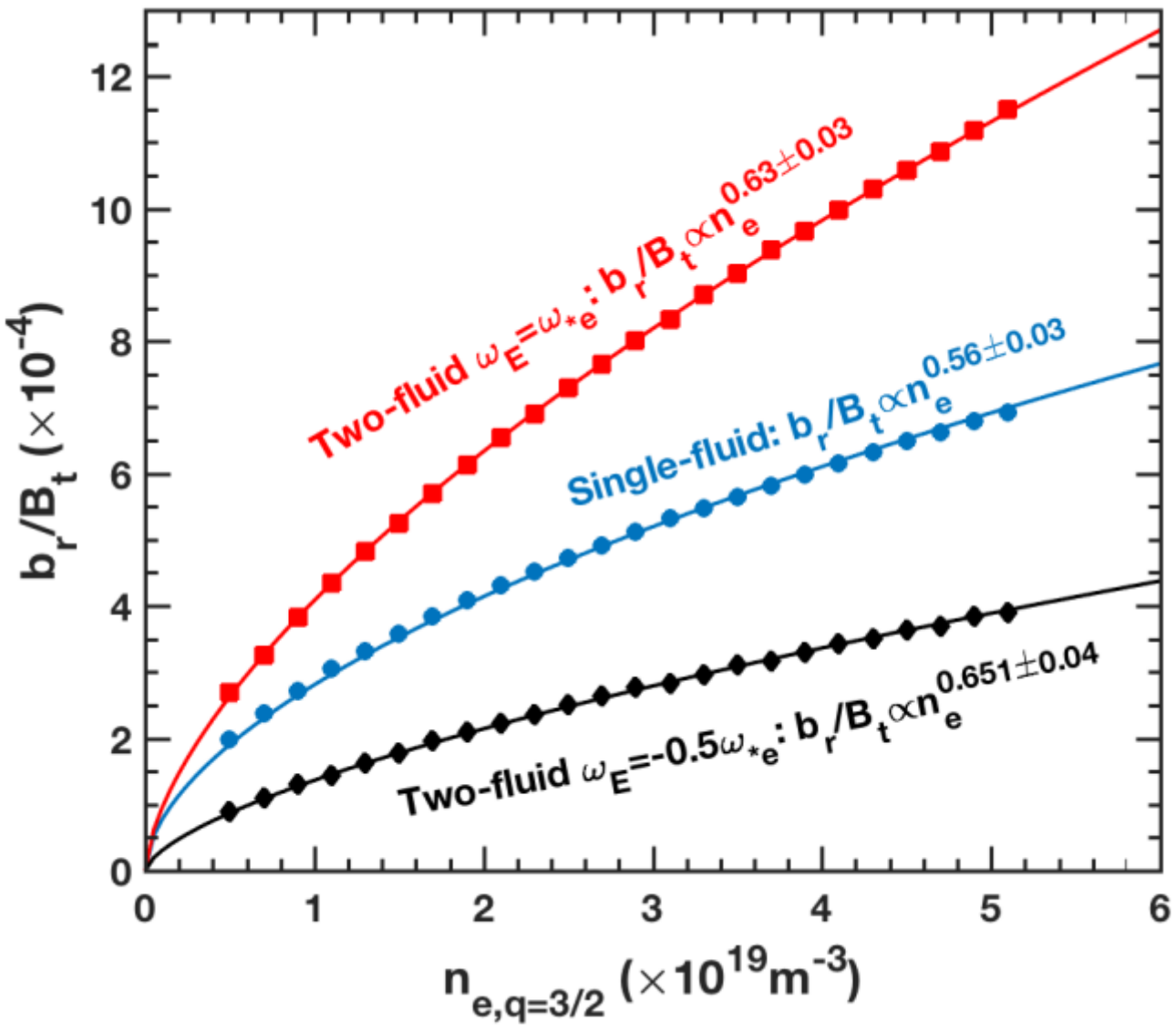

**Figure 8.** Comparison of TM1 single-fluid (blue) and two-fluid simulations of $m/n$ = 3/2 EF penetration threshold versus $n_{e,q=3/2}$ for $\omega_E = \omega_{*e}$ (red) and $\omega_E = -0.5\omega_{*e}$ (black). A least-squares fitting for the numerical results indicates a stronger density scaling of $b_r/B_t \propto n_e^{0.63\pm0.03}$ for $\omega_E = \omega_{*e}$ and $b_r/B_t \propto n_e^{0.651\pm0.04}$ for $\omega_E = -0.5\omega_{*e}$ compared to single-fluid scaling.

The dependence of the EF threshold on density is studied by TM1 two-fluid modeling with $\omega_E = \omega_{*e}$ and $\omega_E = -0.5\omega_{*e}$ as shown in figure 8. In the modeling, the density profile is proportionally changed to scan the density, which leads to a constant $\omega_{*e}$ for different density. Figure 8 shows that the required EF threshold for penetration at $\omega_E = -0.5\omega_{*e}$ ($\omega_{\perp e} = 0.5\omega_{*e}$) is lower than that of single fluid ($\omega_{\perp e} = \omega_E = \omega_{*e}$), while the EF threshold for $\omega_E = \omega_{*e}$ ($\omega_{\perp e} = 2\omega_{*e}$) is higher than that of single fluid. Figure 8 indicates that plasma with higher natural frequency $\omega_{\perp e}$ requires stronger EF for penetration. The two-fluid modeled EF threshold has a stronger dependence on density with $b_r/B_t \propto n_e^{0.63\pm0.03}$ for $\omega_E = \omega_{*e}$ and $b_r/B_t \propto n_e^{0.651\pm0.04}$ for $\omega_E = -0.5\omega_{*e}$ compared to the single-fluid scaling (blue) $b_r/B_t \propto n_e^{0.56\pm0.03}$. The reason is that higher density leads to higher collisionality (and lower $d_1$), which in turn results in weaker parallel transport. Weaker parallel transport leads to less change in electron pressure and less reduction in $\omega_{*e}$, which requires even stronger EF amplitude to penetrate.

The EF scaling on density is further scanned for different $\omega_E/\omega_{*e}$ ranging from -10 to 10 and the modeled $\alpha_n$ is shown as a function of $\omega_E/\omega_{*e}$ in figure 9(a). It is found that, on the one hand, $\alpha_n$ is obviously higher than the single-fluid scaling when the magnitude of $\omega_{*e}$ is comparable to $\omega_E$ with $-3 < \omega_E/\omega_{*e} < 3$. $\alpha_n$ approaches the maximum value when $\omega_E/\omega_{*e} \sim -1$. On the other hand, the two-fluid scaling approaches the single-fluid scaling when $\omega_{*e}$ is relatively negligible compared to plasma rotation $\omega_E$, i.e. $|\omega_E/\omega_{*e}| \sim 10$. It is reasonable to expect the variation of $\alpha_n$ on $\omega_E/\omega_{*e}$ as shown in figure 9(a), since the plasma transport caused by EF is more relevant when $\omega_{*e}$ is comparable to $\omega_E$, and the inverse dependence of parallel transport on collisionality (and hence density) requires additional stronger EF to brake down the rotation frequency contributed by $\omega_{*e}$.

We note that in the DIII-D Ohmic or L-mode discharges, $\omega_E/\omega_{*e}$ usually ranges from -2 to 2, which means the two-fluid effect should be taken into account according to the two-fluid model. It should be noted that distinguishing between 0.56 and 0.67 would be difficult if not impossible with the quality experimental of data currently available due to the difficulty in isolating a single parameter experimentally. In addition, the EF scaling should exhibit a stronger density dependence compared to single-fluid scaling ($\alpha_n \sim 0.5$).

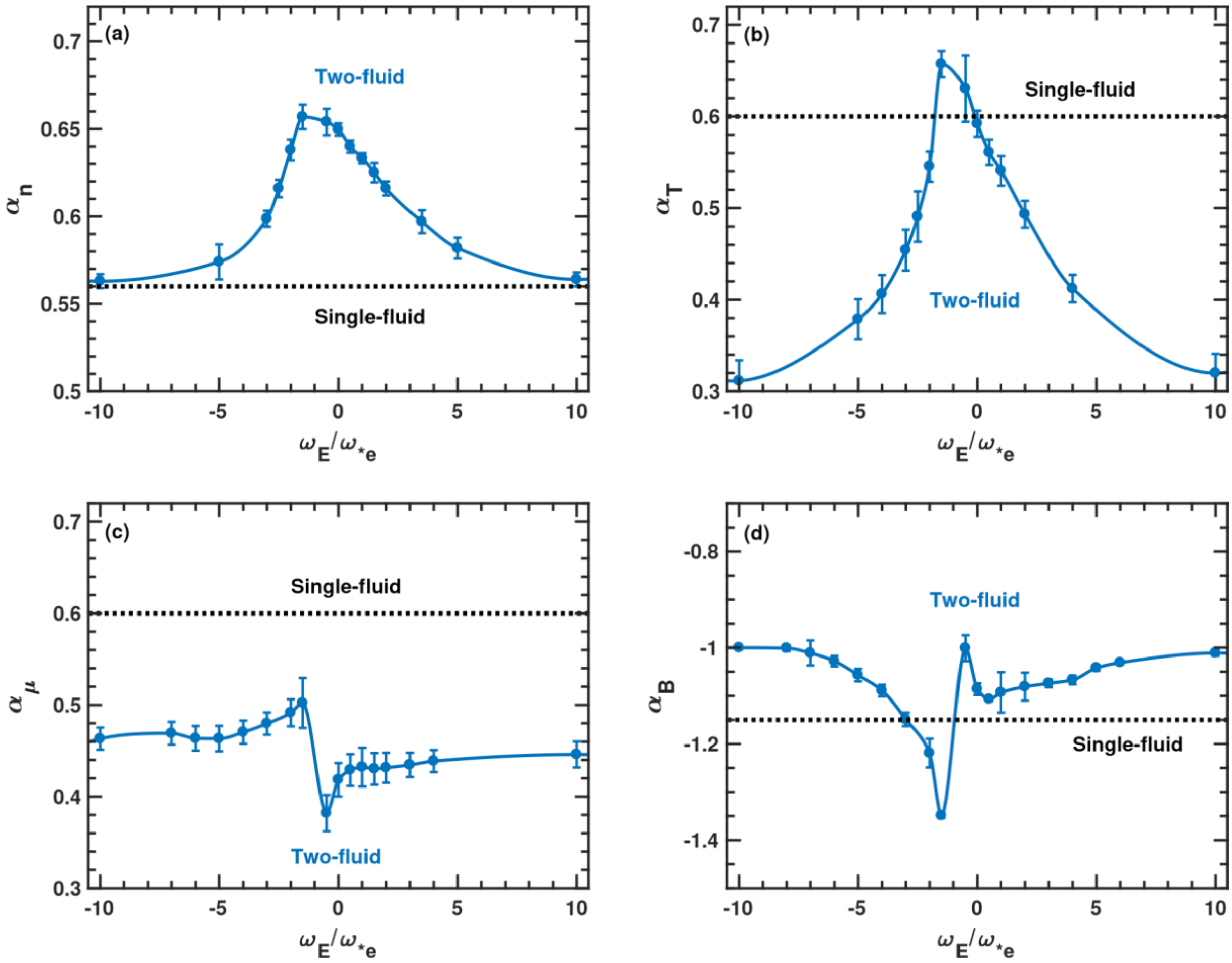

**Figure 9.** TM1 two-fluid simulation of the scaling coefficients (a) $\alpha_n$ on density, (b) $\alpha_T$ on temperature, (c) $\alpha_\mu$ on viscosity and (d) $\alpha_B$ on toroidal field are shown as a function of the ratio between $\omega_E$ and $\omega_{*e}$. Here, $\alpha$ from single-fluid modeling is shown in black dotted line for comparison.

The two-fluid modeled $\alpha_T$ on temperature is shown as a function of $\omega_E/\omega_{*e}$ in figure 9(b). Here, $\omega_{*e}$ is artificially kept constant when scanning $T_e$ to simplify the EF scaling on $T_e$. It is found that, on the one hand, $\alpha_T$ is a little higher than the singlefluid scaling when $-2 < \omega_E/\omega_{*e} < 0$. On the other hand, $\alpha_T$ decreases below the single-fluid scaling with increasing $|\omega_E/\omega_{*e}|$, and it even decreases to near $\alpha_T = 0.3$ when $|\omega_E/\omega_{*e}| = 10$. It is reasonable to expect the variation of $\alpha_T$ with $\omega_E/\omega_{*e}$ as shown in figure 9(b), since on the one hand, $-1 < \omega_{\perp e}/\omega_{*e} < 1$ when $-2 < \omega_E/\omega_{*e} < 0$, the low natural frequency requires weaker EF threshold to penetrate, which leads to a smaller saturated island width and less effect on the particle and energy transport. As a result, the applied EF will accelerate the plasma rotation frequency $\omega_E$ [50] when $-1 < \omega_E/\omega_{*e} < 0$ or increase the electron pressure gradient at the $q = 3/2$ surface [39] when $-2 < \omega_E/\omega_{*e} < -1$ to cause $\omega_{\perp e} = 0$, resulting in a stronger temperature dependence with $\alpha_T > 0.6$. On the other hand, the increasing $|\omega_E/\omega_{*e}|$ leads to higher $|\omega_{\perp e}|$ when $\omega_E/\omega_{*e} < -2$ or $\omega_E/\omega_{*e} > 0$. Higher $|\omega_{\perp e}|$ requires stronger EF to penetrate and form magnetic island with larger width. The larger magnetic island will cause stronger effect on the particle and energy transport due to the positive dependence of parallel transport on the temperature, leading to a weaker dependence of EF threshold on temperature with $\alpha_T < 0.6$. According to the two-fluid scaling shown in figure 9(b), the EF scaling on temperature should be close to the singlefluid scaling for the DIII-D Ohmic or L-mode discharges with $-2 < \omega_E/\omega_{*e} < 2$.

The two-fluid modeled $\alpha_\mu$ on plasma viscosity is shown as a function of $\omega_E/\omega_{*e}$ in figure 9(c). It is found that, $\alpha_\mu$ almost keeps constant with $\alpha_\mu \sim 0.45$ except a jump in the region of $-2 < \omega_E/\omega_{*e} < 0$. In the two-fluid model, on the one hand, the contribution of $\omega_{*e}$ in $\omega_{\perp e}$ lowers the effect of plasma rotation and hence plasma viscosity in determining EF threshold. On the other hand, stronger plasma viscosity leads to more change in the particle and energy transport [39] and hence more change in $\omega_{*e}$, which in turn makes $\omega_{\perp e}$ easier to approach 0. Theses two effects lower the dependence of EF scaling on $\mu$. In addition, the applied EF accelerates the plasma rotation [50] in the region of $-1 < \omega_E/\omega_{*e} < 0$ and leads to a stronger dependence of EF scaling on plasma viscosity, which results in the jump in $\alpha_\mu$ in that region. The results in figure 9(c) reveals that, in the two-fluid model, there is always a dependence of EF scaling on density even in the linear Ohmic confinement region with $\tau_\mu \propto \tau_E \propto n^{-}{}_e{}^1$ according to the Neo-Alcator scaling [27], which is an important advance for EF scaling compared to single-fluid or theoretical scaling.

The two-fluid modeled $\alpha_B$ on toroidal field is shown as a function of $\omega_E/\omega_{*e}$ in figure 9(d). The increasing magnetic field leads to an increasing $\omega_{*e}$, here, $\omega_{*e}$ is artificially kept constant when scanning $B_t$ to simplify the EF scaling. It is found that, $\alpha_B$ decreases slightly from -1.15 to -1 except a jump in the region of $-2 < \omega_E/\omega_{*e} < 0$, indicating two-fluid effects slightly affect the EF scaling on $B_t$.

The dependence of EF threshold on plasma rotation $\omega_E$ is modeled in the two-fluid model, and the saturated island width is shown by the contour plot versus EF amplitude $B_r$ and $\omega_E/\omega_{*e}$ in figure 10(a). The white line in figure 10(a) corresponds to the penetration threshold, which divides the bifurcation from screening (blue) to penetration (yellow). It is found that there is a minimum EF amplitude for field penetration at $\omega_E/\omega_{*e} = -1$, where $\omega_{\perp e} = \omega_E + \omega_{*e} = 0$ due to the cancelling between the equilibrium plasma rotation frequency $\omega_E$ and the diamagnetic drift frequency $\omega_{*e}$. The EF penetration threshold linearly increases as the rotation frequency deviates from the frequency ($\omega_{EF} = 0$) of EF, which can be expressed as $b_r/B_t \propto |\omega_{\perp e}|$. It should be noted that, the scaling becomes $b_r/B_t \propto |\omega_{\perp e} - \omega_{EF}|$ when the EF is rotating in the frequency of $\omega_{EF}$. This linear dependence of EF threshold on rotation frequency is well understood in theory [4] and confirmed by experiment [8, 12, 13, 15].

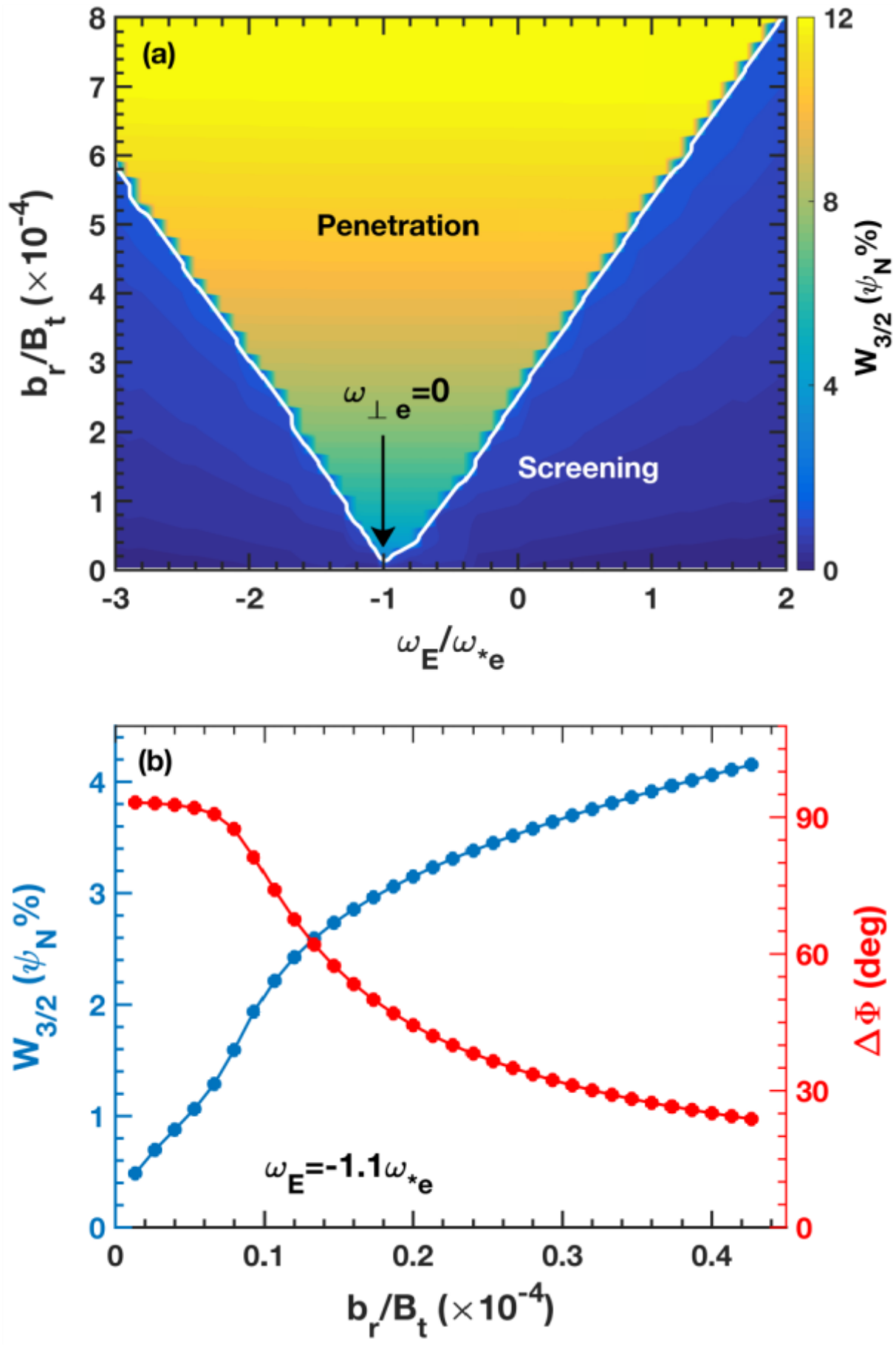

**Figure 10.** (a) TM1 two-fluid simulation of $m/n = 3/2$ EF penetration threshold versus plasma rotation represented by the contour plot of the saturated island width $W_{3/2}$ versus $b_r/B_t$ and $\omega_E/\omega_{*e}$. The penetration threshold is shown by the white curve, which shows a linear dependence of the penetration threshold on rotation. (b) TM1 two-fluid simulated island width $W_{3/2}$ and phase difference $\Delta\Phi$ versus EF amplitude $b_r$ for $\omega_E = -1.1\omega_{*e}$.

An enduring mystery in EF experiments is that the penetration threshold is non-zero [13, 15] even when the plasma natural frequency is zero ($\omega_{\perp e} \sim 0$). From our modeling as shown in figure 10(a), on the one hand, we find that the island width is much smaller when $\omega_{\perp e} \sim 0$ ($\omega_E/\omega_{*e} = -1$) due to the much lower EF penetration threshold. This very small magnetic island makes it very hard to measure the onset of locked mode until it grows up to large enough width to be detected by magnetic measurement. On the other hand, the saturated island almost linearly increases with the increasing EF amplitude, the magnetic perturbation generated by the magnetic island may probably be treated as the pick-up from the EF. Figure 10(b) shows the two-fluid simulated island width $W_{3/2}$ and phase difference $\Delta\Phi$ as a function of $b_r/B_t$ when $\omega_E = -1.1\omega_{*e}$ ($\omega_{\perp e} = -0.1\omega_{*e}$). Different to figure 2(a)-(c) with a fast jump in both $W_{3/2}$ and $\Delta\Phi$ when penetration happens, $W_{3/2}$ and $\Delta\Phi$ continuously vary with the increasing EF amplitude without a bifurcated jump as shown in figure 10(b).

According to the theoretical model (Equations (3-7)) of TM1, it is clear that both the penetration threshold and the resulting transport are sensitive to the transport coefficients at the rational surface. To confirm the sensitivity of the simulated scaling to the choice of transport coefficients, simulations of density scaling are performed with $\omega_E = \omega_{*e}$ for a range of transport coefficients, $\mu = \chi_\perp = 2D_\perp = 0.25\ m^2/s$ (red), 0.5 $m^2/s$ (blue) and 1 $m^2/s$ (black), as shown in figure 11. It is found that smaller transport coefficients decrease the EF penetration threshold, making field penetration easier to achieve, and vice versa for larger transport coefficients. The reason is that, on the one hand, stronger viscosity leads to higher penetration threshold as discussed in figure 5. On the other hand, the ion polarization current caused by the RMP, affects the parallel collisional transport of heat and particle as indicated by the first term in the right side of Equations (6) and (7). The density and temperature will change more (less) for a smaller (larger) transport coefficient, leading to more (less) flattening in $n_e$ and $T_e$ and hence more (less) change in $\omega_{*e}$. However, the EF scaling on density is the same as shown in figure 11, which indicates that the the unified change of transport coefficient won't affect the EF scaling.

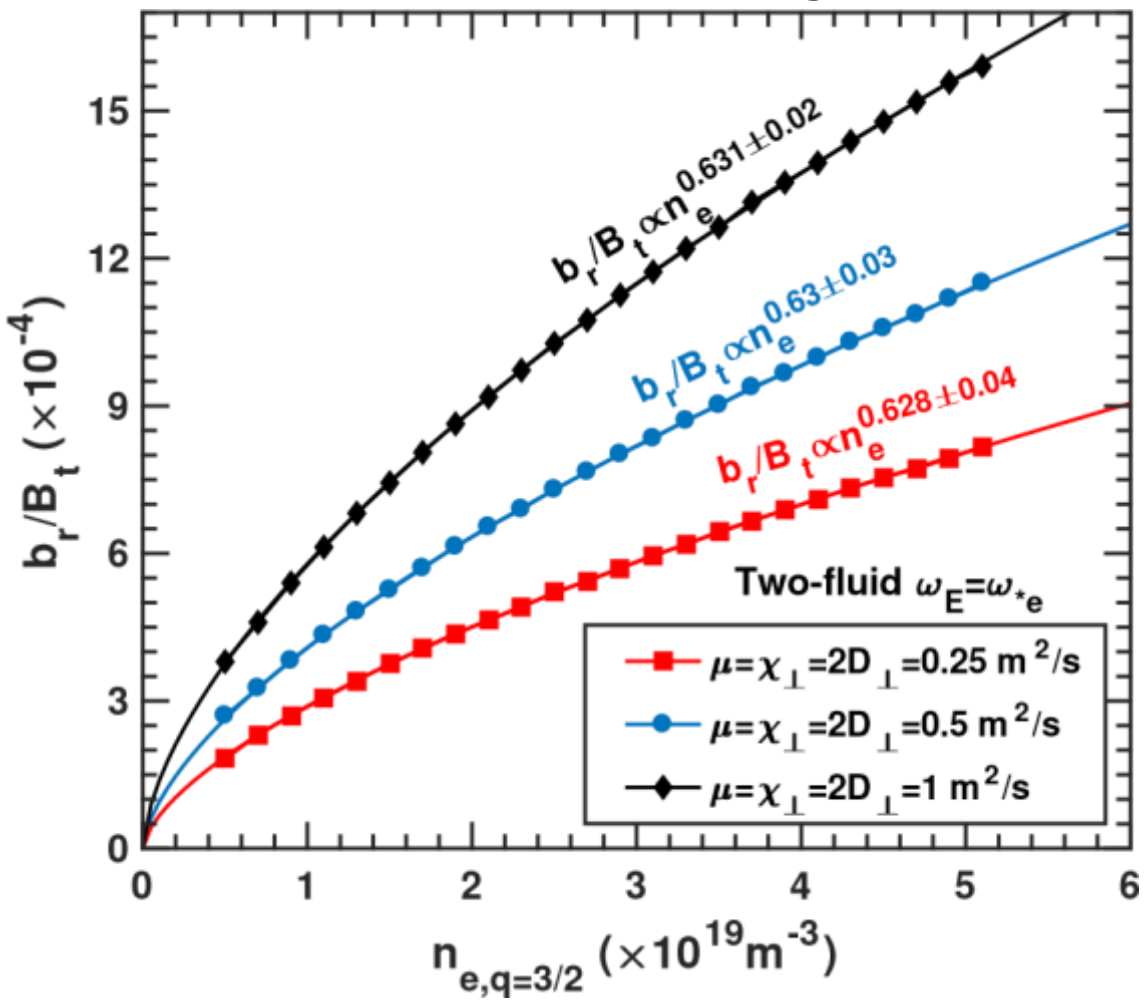

**Figure 11.** Effect of transport coefficients on EF scaling. Comparison of $m/n = 3/2$ EF penetration threshold versus $n_{e,q=3/2}$ for $\omega_E = \omega_{*e}$ (red) with different transport coefficients of $\mu = \chi_\perp = 2D_\perp = 0.25\ m^2/s$ in red, $\mu = \chi_\perp = 2D_\perp = 0.5\ m^2/s$ in blue and $\mu = \chi_\perp = 2D_\perp = 1\ m^2/s$ in balck. A least-squares fitting for the numerical results indicates a very similar density scaling.

It is reasonable to expect non-constant scaling coefficients on plasma parameters from twofluid model as shown in figure 9. Since the twofluid model here we used has spatial dependence, i.e. the partial derivate in both the density and temperature in Equations (3), (6) and (7). These partial derivatives affect the diamagnetic drift frequency $\omega_{*e}$ and the transport in both particle and energy across the resonant surface, and of course they couple together. The former one relates to the mode natural frequency $\omega_{\perp e}$ and can be scaled equivalently by scanning the plasma rotation frequenc $\omega_E$. While the latter one would make the EF scaling much more complicated and is hard to identified due to limitations in the spatial resolution of profile measurements in experiments.

The comparison between two-fluid and singlefluid simulations in figure 9 reveals the similarity in the EF scaling except the former one has: 1) a stronger dependence on density at lower ratio of $\omega_E/\omega_{*e}$; 2) a weaker dependence on plasma viscosity; 3) a weaker dependence on electron temperature at higher ratio of $\omega_E/\omega_{*e}$ due to the larger magnetic island width. In addition, for plasma with higher electron temperature, both the electron diamagnetic drift frequency and parallel collisional transport becomes more important, and two-fluid model will address the EF scaling more accurately
[28–30].

## 4. Discussion and summary

The scaling of $m/n = 3/2$ single helicity EF penetration threshold on plasma parameters is studied in both single-fluid and two-fluid models by using nonlinear MHD modeling with a cylindrical circular tokamak approximation. This cylindrical geometry is different from the DIII-D highly shaped plasma. On the one hand, this approximation neglects the toroidal mode coupling effect, which possibly affects the solution of the mode/magnetic island during the nonlinear stage before the occurrence of EF penetration. However, this nonlinear toroidal mode coupling will be suppressed by the plasma rotation shear (and hence frequency difference between different rational surfaces) with the absence of any modes/magnetic islands [51]. On the other hand, the modifications of plasma response expected from a toroidal plasma ideal MHD mode is also neglected, which will modify resonant fields at the corresponding rational surface [52]. However, this approximation in the TM1 code enables the identification of the pure EF scaling on plasma parameters without additional effect. The application of the full toroidal code GPEC [53] may minimize the weakness of that approximation in TM1, and make it possible to quantitatively compare or predict experiments [43, 54]. In detail, GPEC calculates both the vacuum and ideal response of magnetic perturbations by including all the geometry effects, and the total of these two terms are the actually applied effective EF strength [26]. What TM1 modeled here is the required total EF amplitude for field penetration. Hence, to compare or implement the EF scaling laws to the experiment, we need to compare the TM1 modeled EF threshold and the actual experimental RMP strength calculated by GPEC. By doing so, the ignored toroidal coupling in the penetration threshold and certainly on the saturated island width should be weak, however this issue awaits future study using nonlinear toroidal codes.

It should be pointed out that a reliable EF scaling from both experiment and theory/modeling will benefit the prediction of EF tolerance in ITER. For the experimental empirical scaling, the more relevant parameters (i.e. $T_e$) are included in the scaling, the more reliable it will be. Especially, the coupling of plasma parameters must be also taken into account during the experimental scanning. Considering the high temperatures EF scaling based on nonlinear two-fluid theory or modeling will more accurately.

In addition, the plasma parameters scanned in the modeling correspond to the collisional and semi-collisional regimes [29], which is valid for the modeling by using the MHD code TM1. Due to the wide range of parameters scanned in the scaling, the modeled EF scaling in this paper is not only reliable for the Ohmic or L-mode plasma, it is also reliable to address EF penetration in high performance plasma like H-mode. In the core of H-mode plasma $\omega_E/\omega_{*e} < -2$, accordingly, the EF scaling on density will be very sensitive to the plasma rotation $\omega_E$ with a weaker dependence on the temperature. In the edge of H-mode plasma, the externally applied RMPs are found to cause density pump out and edge localized modes (ELMs) suppression in DIII-D [43, 54–57]. TM1 modeling of the plasma response to RMP finds that field penetration at the top of pedestal that causes ELM suppression scales as $b_r/B_t \propto n_e^{0.7}\omega_{\perp e}^{0.94}$ [54]. The scaling of pedestal top penetration threshold is consistent with the EF scaling described in this paper with a stronger dependence on density, since $0 < \omega_E/\omega_{*e} << 1$ due to the much higher diamagnetic drift frequency $\omega_{*e}$ at the top of pedestal. The dependence of RMP penetration threshold on both density and rotation also consistently explains the observed access conditions with low density and high toroidal rotation for ELM suppression [58, 59]. As a result, a reliable EF scaling can also help to explain and predict the access condition for controlling ELMs by RMP.

The TM1 single-fluid model has been used to estimate the ITER $n$ = 1 and $n$ = 2 EF penetration thresholds using DIII-D ITER Baseline Scenario discharge profiles scaled to the following parameters: $T_e$ = 1 keV and $n_e = 5\times10^{19}m^{-3}$ at the rational surface ($q$ = 2 or $q$ = 3/2 surface) combined with the ITER size ($R_0$ = 6.2 m), toroidal field ($B_t$ = 5.2 T) and $q_{95}$ = 3.3. Estimating the ITER $E \times B$ rotation frequency between 2-8 krad/s. Here, single-fluid instead of two-fluid simulation is performed to avoid the complexity of the coupled two-fluid effects. The TM1 simulation predicts thresholds spanning $b_r/B_t = 0.4 - 1.5 \times 10^{-4}$ for $n$ = 1 and $b_r/B_t = 1 - 3 \times 10^{-4}$ for $n$ = 2 in
ITER, which are lower than the EFs threshold in DIII-D. This is consistent with the simple experimental regression projection [26] and theoretical predictions [28] for $n$ = 1 and consistent with the experience that the $n$ = 2 threshold tends to be on the same order as the $n$ = 1 threshold on current machines .

In summary, the scaling of the EF penetration threshold on plasma parameters is examined by nonlinear MHD modeling from both single-fluid and two-fluid model. The conclusions of this work are:

(i) Nonlinear *single-fluid* modeling finds the EF threshold scales as $b_r/B_t \propto n_e^{0.56} T_e^{0.6} \tau_\mu^{-0.59} B_t^{-1.15} \omega$, which is very similar
to the theoretical scaling in both *Rutherford* and *visco-resistive* regimes [4].

(ii) Nonlinear *two-fluid* modeling finds that the EF scaling on density, temperature, plasma viscosity and toroidal field depends significantly on the ratio of $\omega_E/\omega_{*e}$.

(iii) The EF penetration threshold linearly depends on the perpendicular flow frequency $\omega_{\perp e}$, and it is minimized when $\omega_{\perp e} = \omega_E + \omega_{*e} = 0$. The very small island and smooth bifurcation in EF penetration near zero frequency is hard to detect in the experiment, leading to a finite penetration threshold within the capability of the experimental measurements.

## Acknowledgments

This material is based upon work supported by the U.S. Department of Energy, Office of Science, Office of Fusion Energy Sciences, using the DIII-D National Fusion Facility, a DOE Office of Science user facility, under Awards DE-AC02-09CH11466 and DE-FC02-04ER54698.